\documentclass[a4paper,12pt]{article}

\usepackage{graphics, epsfig}

\begin{document}

\author{C. Barrab\`es\thanks{E-mail : barrabes@lmpt.univ-tours.fr}\\
\small Laboratoire de Math\'ematiques et Physique Th\'eorique,\\
\small  CNRS/UMR 7350, F\'ed\'eration Denis Poisson CNRS/FR2964,\\
\small Universit\'e F. Rabelais, 37200 Tours,
France\\and\\P.A. Hogan\thanks{E-mail : peter.hogan@ucd.ie}\\
\small School of Physics,\\
\small University College Dublin, Belfield, Dublin 4, Ireland}

\title{Colliding Impulsive Gravitational Waves and a Cosmological Constant}
\date{PACS numbers: 04.40.Nr, 04.30.Nk}
\maketitle

\begin{abstract}
We present a space--time model of the collision of two homogeneous, plane impulsive gravitational waves (each having a delta function profile) 
propagating in a vacuum before collision and for which the post collision space--time has constant curvature. The profiles of the incoming waves are $k\,\delta(u)$ 
and $l\,\delta(v)$ where $k, l$ are real constants and $u=0, v=0$ are intersecting null hypersurfaces. The cosmological constant $\Lambda$ in the post collision region 
of the space--time is given by $\Lambda=-6\,k\,l$. \end{abstract}
\thispagestyle{empty}
\newpage

\section{Introduction}
\setcounter{equation}{0}
\noindent
Finding the space--time structure after the collision of gravitational and/or electromagnetic waves is a difficult problem in general relativity due to the non--linearity of the 
field equations. The problem is simplified by specialising to impulsive and/or shock waves which are plane and homogeneous and then exact solutions can be found, with 
the Khan--Penrose \cite{KP} and Bell--Szekeres \cite{BS} solutions among the most famous. Up to now no solution where a cosmological constant appears after the 
collision of two homogeneous, plane, impulsive gravitational waves has yet been found. This paper provides such an exact solution.

Impulsive light--like signals, i.e. signals travelling with the speed of light and having a delta function profile, are idealised models of more realistic light--like signals having a 
profile with a finite width such as, for instance, a burst of light--like matter or of radiation. In some situations they may provide solvable models to describe their interactions and are 
sometimes used as classical models of quantum phenomena. In black hole physics one has the examples of mass inflation \cite{PoI}, the limiting curvature principle \cite{Mark}, 
Hawking radiation and quantum fluctuations \cite{York} and internal structure of a Schwarzschild black hole \cite{FMM}, \cite{FB}. In general relativity an impulsive light--like 
signal exists whenever the Riemann curvature tensor of the space--time manifold exhibits a delta function term with support on a null hypersurface, with the latter representing the 
space--time history of the signal and across which the first derivatives of the metric tensor are discontinuous. This signal can be a thin shell of light--like matter, or an impulsive 
gravitational wave or a mixture of both \cite{BH1}. Recently the cosmological constant has received much attention in connection with a possible description of dark energy. 
Such exotic matter is described by a perfect fluid with an equation of state for which the sum of the energy density and isotropic pressure vanishes. In this paper we examine the 
possibility that the collision of two impulsive gravitational waves will produce such exotic matter by adopting a mathematical point of view, i.e. by solving Einstein's field equations with 
appropriate boundary conditions. No attempt is made to propose a physical mechanism.

For most of the known wave collision models in general relativity the same field equations apply before and after the collision. However the choice of field equations before collision does not 
determine the choice of field equations after collision. This freedom offers an opportunity to explore new and potentially interesting models. The well known space--time model of a head--on collision of two homogeneous, plane impulsive gravitational waves, travelling in a vacuum, involves the assumption 
that \emph{the post collision region of space--time is a vacuum space--time}. With this assumption the post collision region is described by the Khan--Penrose \cite{KP} 
solution of Einstein's vacuum field equations (for a derivation see \cite{BH}). We demonstrate here that if the post collision region of space--time is assumed to be a solution 
of \emph{Einstein's field equations with a cosmological constant} then an exact solution of these field equations can be found satisfying the same conditions on the null hypersurface boundaries 
of the post collision region as the Khan--Penrose solution. In addition the cosmological constant can be expressed simply in terms of two parameters which label each of the incoming 
waves. The post collision region is a space--time of constant curvature and is thus curvature singularity--free, in contrast to the Khan--Penrose model. The solution derived here 
is not an extension of the Khan--Penrose solution since it has the property that if the cosmological constant vanishes then at least one of the incoming waves vanishes. The post collision 
model presented here can be explained in terms of a redistribution of the energy in the incoming waves and this is described in some detail.

In section 2 the incoming plane, impulsive gravitational waves propagating through a vacuum are introduced, the collision problem is specified (as a light--like boundary value problem) 
and the solution of Einstein's field equations with a cosmological constant in the post collision region is given. This is followed in section 3 by a detailed study of the physical 
properties of the products of the collision which, in addition to a cosmological constant, include impulsive gravitational waves (as in the Khan--Penrose collision) and light--like shells of 
matter. When reasonable physical restrictions are invoked the post collision region of space--time is anti-- de Sitter space--time in this case.

\section{Colliding Waves}
\setcounter{equation}{0}
\noindent
A plane, homogeneous gravitational impulse wave propagating in a vacuum is described in general relativity by a space--time with line element
\begin{equation}\label{1}
ds^2=-(1+k\,u_+)^2dx^2-(1-k\,u_+)^2dy^2+2\,du\,dv\ ,\end{equation}
where $k$ is a constant (introduced for convenience) and $u_+=u\,\vartheta(u)$ where $\vartheta(u)=1$ for $u>0$ and $\vartheta(u)=0$ for $u<0$ is 
the Heaviside step function. The metric given via this line element satisfies Einstein's vacuum field equations everywhere (in particular on $u=0$). The only 
non--vanishing Newman--Penrose component of the Riemann curvature tensor on the tetrad given via the 1--forms $\vartheta^1=(1+k\,u_+)dx\ ,\ 
\vartheta^2=(1-k\,u_+)dy\ ,\ \vartheta^3=dv\ ,\ \vartheta^4=du$ is
\begin{equation}\label{2}
\Psi_4=-k\,\delta(u)\ .\end{equation}Thus the curvature tensor is type N (the radiative type) in the Petrov classification with the vector field $\partial/\partial v$ the degenerate 
principal null direction and therefore the propagation direction of the history of the wave (the null hypersurface $u=0$) in space--time. The wave profile is the delta function, 
singular on $u=0$, and thus the wave is an impulsive wave. There are two families of intersecting null hypersurfaces $u={\rm constant}$ and $v={\rm constant}$ in the space--time 
with line element (\ref{1}). A homogeneous, plane impulsive gravitational wave propagating in a vacuum in the opposite direction to that with history $u=0$ has history $v=0$ and 
this is described by a space--time with line element
\begin{equation}\label{3}
ds^2=-(1+l\,v_+)^2dx^2-(1-l\,v_+)^2dy^2+2\,du\,dv\ ,\end{equation}
where $l$ is a convenient constant and $v_+=v\,\vartheta(v)$. The Ricci tensor vanishes everywhere when calculated with the metric tensor given by this line element. The only 
non--vanishing Newman--Penrose component of the Riemann curvature tensor on the tetrad given via the 1--forms $\vartheta^1=(1+l\,v_+)dx\ ,\ 
\vartheta^2=(1-l\,v_+)dy\ ,\ \vartheta^3=dv\ ,\ \vartheta^4=du$ is
\begin{equation}\label{4}
\Psi_0=-l\,\delta(v)\ ,\end{equation}indicating a Petrov type N curvature tensor with degenerate principal null direction $\partial/\partial u$.

For the collision problem we envisage a pre--collision vacuum region of space--time $v<0$ with line element (\ref{1}) and a pre--collision vacuum region of space--time $u<0$ with 
line element (\ref{3}) (with both line elements coinciding when $v<0$ \emph{and} $u<0$). The waves collide at $u=v=0$ and the post collision region of the space--time corresponds 
to $u>0$ \emph{and} $v>0$. In this region the line element has the form (\cite{KP}, \cite{S1}, \cite{S2})
\begin{equation}\label{5}
ds^2=-e^{-U}(e^Vdx^2+e^{-V}dy^2)+2\,e^{-M}du\,dv\ ,\end{equation}
where $U, V, M$ are each functions of $u, v$. These functions must satisfy the following conditions on the null hypersurface boundaries of the region $u>0\ ,\ v>0$: 
\begin{equation}\label{6}
v=0\ ,\ u\geq0\ \ \Rightarrow\ \ e^{-U}=1-k^2u^2\ ,\ e^V=\frac{1+k\,u}{1-k\,u}\ ,\ M=0\ ,\end{equation}
and
\begin{equation}\label{7}
u=0\ ,\ v\geq0\ \ \Rightarrow\ \ e^{-U}=1-l^2v^2\ ,\ e^V=\frac{1+l\,v}{1-l\,v}\ ,\ M=0\ .\end{equation}Einstein's field equations with a cosmological constant $\Lambda$ in the region $u>0, v>0$ calculated with 
the metric tensor given by the line element (\ref{5}) read:
\begin{eqnarray}
U_{uv}&=&U_u\,U_v-\Lambda\,e^{-M}\ ,\label{a8}\\
2\,V_{uv}&=&U_u\,V_v+U_v\,V_u\ ,\label{a9}\\
2\,U_{uu}&=&U_u^2+V_u^2-2\,M_u\,U_u\ ,\label{a10}\\
2\,U_{vv}&=&U_v^2+V_v^2-2\,M_v\,U_v\ ,\label{a11}\\
2\,M_{uv}&=&V_u\,V_v-U_u\,U_v\ ,\label{a12}\end{eqnarray}where the subscripts denote partial derivatives. To implement our strategy below for solving (\ref{a8})--(\ref{a12}) subject to the 
boundary conditions (\ref{6}) and (\ref{7}) we will need to know $V_v$ at $v=0$, which we denote by $(V_v)_{v=0}$, and $V_u$ at $u=0$, which we denote by $(V_u)_{u=0}$. We already have from (\ref{6}) and (\ref{7}):
\begin{equation}\label{a13}
(V_u)_{v=0}=\frac{2\,k}{1-k^2u^2}\ \ \ {\rm and}\ \ \ (V_v)_{u=0}=\frac{2\,l}{1-l^2v^2}\ ,\end{equation}
and also
\begin{equation}\label{a14}
(U_u)_{v=0}=\frac{2\,k^2u}{1-k^2u^2}\ \ \ {\rm and}\ \ \ (U_v)_{u=0}=\frac{2\,l^2v}{1-l^2v^2}\ .\end{equation}In order to compute $(V_v)_{v=0}$ and $(V_u)_{u=0}$ we must first calculate 
$(U_v)_{v=0}$ and $(U_u)_{u=0}$. We obtain these latter quantities by evaluating (\ref{a8}) at $u=0$ and at $v=0$ and solving the resulting first order ordinary differential equations. The 
constants of integration which arise are determined from the fact that $U_v$ and $U_u$ both vanish when $u=0$ \emph{and} $v=0$, which follows from (\ref{a14}). We then find that 
\begin{equation}\label{a15}
(U_v)_{v=0}=-\frac{\Lambda\,u\,(1-\frac{1}{3}k^2u^2)}{1-k^2u^2}\ \ \ {\rm and}\ \ \ (U_u)_{u=0}=-\frac{\Lambda\,v\,(1-\frac{1}{3}l^2v^2)}{1-l^2v^2}\ .\end{equation}
Now evaluating (\ref{a9}) at $v=0$ and at $u=0$ provides us with a pair of first order ordinary differential equations for $(V_v)_{v=0}$ and $(V_u)_{u=0}$. These equations are 
straightforward to solve and the resulting constants of integration are determined from the fact that $V_u=2\,k$ and $V_v=2\,l$ when $u=0$ \emph{and} $v=0$, which follows from 
(\ref{a13}). The final results are:
\begin{eqnarray}
(V_v)_{v=0}&=&\left (2\,l+\frac{\Lambda}{3\,k}\right )(1-k^2u^2)^{-1/2}-\frac{\Lambda}{3\,k}\left (\frac{1+k^2u^2}{1-k^2u^2}\right )\ ,\label{a16}\\
(V_u)_{u=0}&=&\left (2\,k+\frac{\Lambda}{3\,l}\right )(1-l^2v^2)^{-1/2}-\frac{\Lambda}{3\,l}\left (\frac{1+l^2v^2}{1-l^2v^2}\right )\ .\label{a17}\end{eqnarray}

Dividing (\ref{a9}) successively by $V_u$ and by $V_v$ and then differentiating the resulting equations and combining them we obtain
\begin{equation}\label{a18}
2\,\frac{\partial^2}{\partial u\partial v}\log\frac{V_u}{V_v}=\left (U_u\,\frac{V_v}{V_u}\right )_u-\left (U_v\,\frac{V_u}{V_v}\right )_v\ .\end{equation}
This suggests that we examine the possibility of a separation of variables:
\begin{equation}\label{a19}
\frac{V_u}{V_v}=\frac{A(u)}{B(v)}\ ,\end{equation}
for some functions $A(u)$ and $B(v)$. The resulting mathematical simplification is that (\ref{a19}) becomes a first order wave equation for $V$ (see below) and that (\ref{a18}) 
becomes a second order wave equation for $U$. From a physical point of view we have shown \cite{BH1} that if, as is the case in general, two systems of backscattered 
gravitational waves exist in the post collision region (one with propagation direction $\partial/\partial u$ in space--time and one with propagation direction $\partial/\partial v$) then 
(\ref{a19}) implies that there exists a frame of reference in which the energy densities of the two systems of waves are equal.  Using (\ref{a13}), (\ref{a16}) and (\ref{a17}) determines 
the right hand side of (\ref{a19}) and the result is
\begin{equation}\label{a20}
\frac{V_u}{V_v}=\frac{k\left [\left (1+\frac{\Lambda}{6\,k\,l}\right )\sqrt{1-l^2v^2}-\frac{\Lambda}{6\,k\,l}(1+l^2v^2)\right ]}{l\left [\left (1+\frac{\Lambda}{6\,k\,l}\right )\sqrt{1-k^2u^2}-\frac{\Lambda}{6\,k\,l}(1+k^2u^2)\right ]}\ .\end{equation}
Hence this equation can be written as a first order wave equation
\begin{equation}\label{a21}
V_{\bar u}=V_{\bar v}\ ,\end{equation}
with $\bar u(u)$ and $\bar v(v)$ given by the differential equations
\begin{eqnarray}
\frac{d\bar u}{du}&=&k\left [\left (1+\frac{\Lambda}{6\,k\,l}\right )\sqrt{1-k^2u^2}-\frac{\Lambda}{6\,k\,l}(1+k^2u^2)\right ]^{-1}\ ,\label{a22}\\
\frac{d\bar v}{dv}&=&l\left [\left (1+\frac{\Lambda}{6\,k\,l}\right )\sqrt{1-l^2v^2}-\frac{\Lambda}{6\,k\,l}(1+l^2v^2)\right ]^{-1}\ .\label{a23}\end{eqnarray}
These two equations are interesting in general but we shall concentrate in this paper on two stand--out special cases: $\Lambda=0$ and $\Lambda=-6\,k\,l$. The case $\Lambda=0$ 
is shown in the Appendix to correspond to the Khan--Penrose \cite{KP} space--time.

With $\Lambda=-6\,k\,l$ we can solve (\ref{a22}) and (\ref{a23}), requiring $\bar u=0$ when $u=0$ and $\bar v=0$ when $v=0$, with
\begin{equation}\label{a24}
\bar u=\tan^{-1}k\,u\ ,\ \ \bar v=\tan^{-1}l\,v\ .\end{equation}
By (\ref{a21}) we have $V=V(\bar u+\bar v)$ and the boundary condition (\ref{6}) written in terms of $\bar u, \bar v$ reads: when $\bar v=0,\ V=\log\left (\frac{1+\tan\bar u}{1-\tan\bar u}\right )$. 
Hence
\begin{equation}\label{a25}
V(\bar u+\bar v)=\log\left (\frac{1+\tan(\bar u+\bar v)}{1-\tan(\bar u+\bar v)}\right )\ ,\end{equation}
and restoring the coordinates $u, v$ we have
\begin{equation}\label{a26}
V(u, v)=\log\left (\frac{1-k\,l\,u\,v+k\,u+l\,v}{1-k\,l\,u\,v-k\,u-l\,v}\right )\ ,\end{equation}
for $u\geq0, v\geq0$ provided $\Lambda=-6\,k\,l$. Next writing (\ref{a9}) in terms of the variables $\bar u, \bar v$ and using (\ref{a21}) and (\ref{a25}) we have
\begin{equation}\label{a27}
U_{\bar u}+U_{\bar v}=\frac{8\,\tan(\bar u+\bar v)}{1-\tan^2(\bar u+\bar v)}\ ,\end{equation}
which is easily integrated to yield
\begin{equation}\label{a28}
e^{-U}=C(\bar u-\bar v)\left (\frac{1-\tan^2(\bar u+\bar v)}{1+\tan^2(\bar u+\bar v)}\right )\ ,\end{equation}
where $C(\bar u-\bar v)$ is a function of integration. When $\bar v=0$ the boundary condition (\ref{6}) requires $e^{-U}=1-\tan^2\bar u$ and so $C(\bar u)=1+\tan^2\bar u$. Hence 
restoring the coordinates $u, v$ we have $U(u, v)$ given by
\begin{equation}\label{a29}
e^{-U}=\frac{(1-k\,l\,u\,v)^2-(k\,u+l\,v)^2}{(1+k\,l\,u\,v)^2}\ ,\end{equation}
for $u\geq0, v\geq0$. In the light of  (\ref{a28}) we see that $U$ is a linear combination of a function of $\bar u-\bar v$ and a function of $\bar u+\bar v$ and thus satisfies 
the second order wave equation $U_{\bar u\bar u}=U_{\bar v\bar v}$. This wave equation is the equation that (\ref{a18}) reduces to when (\ref{a21}) holds and the barred coordinates are used.

With $V(u, v)$ and $U(u, v)$ given by (\ref{a26}) and (\ref{a29}) we use the field equation (\ref{a8}) with $\Lambda=-6\,k\,l$ to calculate $M(u, v)$. The result is
\begin{equation}\label{a30}
M(u, v)=2\,\log(1+k\,l\,uv)\ ,\end{equation}
and this clearly satisfies the boundary conditions (\ref{6}) and (\ref{7}). Now with $V, U$ and $M$ determined a lengthy calculation verifies that the remaining field equations 
(\ref{a10})--(\ref{a12}) are automatically satisfied. Thus the line element (\ref{5}) of the post collision region reads
\begin{equation}\label{11}
ds^2=\frac{-(1-k\,l\,u\,v+k\,u+l\,v)^2dx^2-(1-k\,l\,u\,v-k\,u-l\,v)^2dy^2+2\,du\,dv}{(1+k\,l\,u\,v)^2}\ .\end{equation} If in the metric tensor components here we replace $u, v$ by 
$u_+=u\,\vartheta(u), v_+=v\,\vartheta(v)$ we obtain in a single line element the expressions (\ref{1}) and (\ref{3}) for the pre--collision regions and (\ref{11}) for the post collision region. In particular this 
will enable us to calculate the physical properties of the boundaries $v=0\ ,\ u\geq0$ and $u=0\ ,\ v\geq0$ of the post collision region. 

\section{Post Collision Physical Properties}
\setcounter{equation}{0}
\noindent
With our sign conventions, choice of units for which $c=G=1$, and energy--momentum--stress tensor $T_{ab}$, Einstein's field 
equations with a cosmological constant $\Lambda$ read
\begin{equation}\label{12'}
R_{ab}=\Lambda\,g_{ab}-8\,\pi(T_{ab}-\frac{1}{2}T^c{}_c\,g_{ab})\ .\end{equation}Also for a perfect fluid with proper density $\rho$ and isotropic pressure $p$ we have
\begin{equation}\label{12''}
T_{ab}=(\rho+p)\,u_a\,u_b-p\,g_{ab}\ ,\end{equation}where $u_a$, satisfying $u_a\,u^a=1$, is the 4--velocity of a fluid particle. We note that for the exotic matter mentioned 
in section 1, $\rho+p=0$ and thus $T_{ab}=\rho\,g_{ab}$ and, with our sign conventions, Einstein's field equations with this energy--momentum--stress tensor are equivalent to the 
field equations with a cosmological constant $\Lambda=8\pi\rho$.

On the half null tetrad $\vartheta^1=e^{-\frac{1}{2}(U-V)}dx\ ,\ 
\vartheta^2=e^{-\frac{1}{2}(U+V)}dy\ ,\ \vartheta^3=e^{-\frac{1}{2}M}dv\ ,\ \vartheta^4=e^{-\frac{1}{2}M}du$ with $V, U, M$ given by (\ref{a26}), (\ref{a29}) and (\ref{a30}) with $u, v$ replaced by $u_+, v_+$ the 
Ricci tensor components of the space--time are given by
\begin{eqnarray}
R_{ab}&=&-6\,k\,l\,\vartheta(u)\,\vartheta(v)\,g_{ab}+\frac{2\,k\,l\,u_+(k^2u_+^2-3)}{1-k^2u_+^2}\delta(v)\,\delta^3_a\delta^3_b\nonumber\\
&&+\frac{2\,k\,l\,v_+(l^2v_+^2-3)}{1-l^2v_+^2}\delta(u)\,\delta^4_a\delta^4_b\ .\label{12}\end{eqnarray}This confirms that the space--time region $u>0, v>0$ is a 
solution of the field equations with a cosmological constant, $R_{ab}=\Lambda\,g_{ab}$, with $\Lambda=-6\,k\,l$ and that there are light--like shells with the 
boundaries $v=0, u\geq0$ and $u=0, v\geq0$ as histories, corresponding to the delta function terms in (\ref{12}). Here $g_{ab}$ are the (constant) metric tensor components 
on the half null tetrad given via the basis 1--forms $\{\vartheta^a\}$.  The light--like shells have no isotropic surface pressure \cite{BI} and the surface energy densities are $\mu_{(1)}$ and $\mu_{(2)}$ 
given by
\begin{equation}\label{12a}
8\,\pi\,\mu_{(1)}=\frac{\Lambda\,u}{3}\left (\frac{k^2u^2-3}{1-k^2u^2}\right )\ \ \ {\rm on}\ \ v=0\ ,\ \ u\geq0\ ,\end{equation}
and
\begin{equation}\label{12b}
8\,\pi\,\mu_{(2)}=\frac{\Lambda\,v}{3}\left (\frac{l^2v^2-3}{1-l^2v^2}\right )\ \ \ {\rm on}\ \ u=0\ ,\ \ v\geq0\ .\end{equation}
The light--like shells must have positive surface energy densities. The only 
way to realise this on $v=0, u\geq0$ (respectively on $u=0, v\geq0$) is to have $kl>0$ and $k^2u^2<1$ (respectively $kl>0$ and $l^2v^2<1$). Thus the cosmological constant $\Lambda=-6\,k\,l$ 
must be negative. These restrictions on the coordinates are less restrictive than the condition $k^2u^2+l^2v^2<1$ for $u\geq0$ and 
$v\geq0$ required in the Khan--Penrose post collision space--time on account of the presence of the curvature singularity. These restrictions on the coordinates also avoid infinite surface energy 
densities in the shells which are arguably as serious as a curvature singularity. Light--like shells did not appear in the Khan--Penrose model and their presence here is due to the non--zero cosmological 
constant.

This result implies that the energy density of the exotic matter is negative and that consequently only the so--called strong energy condition \cite{HE}, namely, $\rho+p\geq0$ and $\rho+3\,p\geq0$, 
can be satisfied.

The Newman--Penrose components of the Weyl conformal 
curvature tensor are given by 
\begin{equation}\label{13}
\Psi_0=-\frac{l(1+k^2u_+^2)}{1-k^2u_+^2}\delta(v)\ ,\ \ \Psi_4=-\frac{k(1+l^2v_+^2)}{1-l^2v_+^2}\delta(u)\ ,\ \ \Psi_1=\Psi_2=\Psi_3=0\ .\end{equation}
Thus the boundaries $v=0, 0\leq k^2u^2<1$ and $u=0, 0\leq l^2v^2<1$ are the histories of impulsive gravitational waves corresponding to the delta function terms here. 
The post collision region $u>0, v>0$ is conformally flat and is a space--time of constant curvature with Riemann curvature tensor components given by
\begin{equation}\label{14}
R_{abcd}=-2\,k\,l(g_{ad}\,g_{bc}-g_{ac}\,g_{bd})\ .\end{equation}Hence this region of space--time does not possess a curvature singularity, in striking contrast to the 
post collision region of the Khan--Penrose space--time.
\section{Summary}
\setcounter{equation}{0}
\noindent
We can briefly summarise our results as follows: for this model collision the energy in the incoming impulsive gravitational waves is re-distributed after the collision into two light--like shells of matter and 
two impulsive gravitational waves moving away from each other followed by a space--time of constant curvature. When the surface energy densities of the post collision light--like shells of 
matter are required to be positive the space--time of constant curvature must be anti de Sitter space--time.

\begin{appendix}
\setcounter{equation}{0}
\section{The case $\Lambda=0$}\indent
With $\Lambda=0$ we can solve (\ref{a22}) and (\ref{a23}), requiring $\bar u=0$ when $u=0$ and $\bar v=0$ when $v=0$, with
\begin{equation}\label{A1}
\bar u=\sin^{-1}k\,u\ ,\ \ \bar v=\sin^{-1}l\,v\ .\end{equation}
By (\ref{a21}) we have $V=V(\bar u+\bar v)$ and the boundary condition (\ref{6}) written in terms of $\bar u, \bar v$ reads: when $\bar v=0,\ V=\log\left (\frac{1+\sin\bar u}{1-\sin\bar u}\right )$. 
Hence
\begin{equation}\label{A2}
V(\bar u+\bar v)=\log\left (\frac{1+\sin(\bar u+\bar v)}{1-\sin(\bar u+\bar v)}\right )=\log\left [\left (\frac{\cos\bar u+\sin\bar v}{\cos\bar u-\sin\bar v}\right )\left (\frac{\cos\bar v+\sin\bar u}{\cos\bar v-\sin\bar u}\right )\right ]\ .\end{equation}
Restoring the coordinates $u, v$ we have
\begin{equation}\label{A3}
V(u, v)=\log\left [\left (\frac{\sqrt{1-k^2u^2}+l\,v}{\sqrt{1-k^2u^2}-l\,v}\right )\left (\frac{\sqrt{1-l^2v^2}+k\,u}{\sqrt{1-l^2v^2}-k\,u}\right )\right ]\ ,\end{equation}
for $u\geq0, v\geq0$ provided $\Lambda=0$. Writing (\ref{a9}) in terms of the variables $\bar u, \bar v$ and using (\ref{a21}) and (\ref{A2}) we have
\begin{equation}\label{A4}
U_{\bar u}+U_{\bar v}=2\,\tan(\bar u+\bar v)\ .\end{equation}Integrating this results in 
\begin{equation}\label{A5}
e^{-U}=D(\bar u-\bar v)\,\cos(\bar u+\bar v)\ ,\end{equation}
where $D(\bar u-\bar v)$ is a function of integration. With $\bar v=0$ the boundary condition (\ref{6}) requires $e^{-U}=\cos^2\bar u$ and so $D(\bar u)=\cos\bar u$. Restoring the coordinates $u, v$ we 
have $U(u, v)$ given by
\begin{equation}\label{A6}
e^{-U}=\cos(\bar u-\bar v)\,\cos(\bar u+\bar v)=1-k^2u^2-l^2v^2\ .\end{equation}
Now (\ref{a8}) with $\Lambda=0$ is automatically satisfied. Combining (\ref{a8}) with $\Lambda=0$ and (\ref{a12}) we have
\begin{equation}\label{A7}
(2\,M+U)_{uv}=V_u\,V_v\ .\end{equation}
Changing the independent variables $u, v$ to $\bar u, \bar v$ using (\ref{A1}), and using (\ref{A2}), this reads
\begin{equation}\label{A8}
(2\,M+U)_{\bar u\bar v}=4\,\sec^2(\bar u+\bar v)\ .\end{equation}Integrating and using (\ref{A6}) we arrive at
\begin{equation}\label{A9}
e^{-2\,M}=\frac{\cos^3(\bar u+\bar v)}{\cos(\bar u-\bar v)}f(\bar u)\,g(\bar v)\ ,\end{equation}
where $f(\bar u)$ and $g(\bar v)$ are functions of integration. From the boundary conditions (\ref{6}) and (\ref{7}) we see that $M=0$ when $\bar v=0$ and $M=0$ when $\bar u=0$ 
and so it follows that $f(\bar u)\,g(\bar v)=\sec^2\bar u\,\sec^2\bar v$. Hence
\begin{equation}\label{A10}
e^{-M}=\sqrt{\frac{\cos^3(\bar u+\bar v)}{\cos(\bar u-\bar v)}}\sec\bar u\,\sec\bar v\ .\end{equation}Restoring the coordinates $u, v$ and simplifying this becomes
\begin{equation}\label{A11}
e^{-M}=\frac{(1-k^2u^2-l^2v^2)^{3/2}}{(\sqrt{1-k^2u^2}\sqrt{1-l^2v^2}+k\,l\,u\,v)^2\sqrt{1-k^2u^2}\sqrt{1-l^2v^2}}\ ,\end{equation}
for $u\geq0, u\geq0$ when $\Lambda=0$. Substituting (\ref{A3}), (\ref{A6}) and (\ref{A11}) into the field equations (\ref{a10}) and (\ref{a11}) verifies that these latter equations are automatically 
satisfied. The line element (\ref{5}) with $V, U$ and $M$ given by (\ref{A3}), (\ref{A6}) and (\ref{A11}) is the Khan--Penrose post collision line element. No derivation is given in \cite{KP}.

\end{appendix}

\end{document}